\documentclass[12pt]{iopart}
\usepackage{iopams}  
\usepackage{xstring} 
\usepackage{hyperref}
\hypersetup{colorlinks=true}
\usepackage{color}
\usepackage{lineno}
\usepackage{aas_macros} 
\usepackage{multirow,bigdelim,makecell,booktabs}
\usepackage{xargs} 

\usepackage[utf8]{inputenc}
\usepackage{tikz}
\usetikzlibrary{positioning}
\usetikzlibrary{shapes.geometric}
\usetikzlibrary{shapes.multipart}
\usetikzlibrary{shapes.misc} 

\usepackage[colorinlistoftodos,prependcaption,textsize=scriptsize]{todonotes}
\newcommandx{\unsure}[2][1=]{\todo[linecolor=red,backgroundcolor=red!25,bordercolor=red,#1]{#2}}
\newcommandx{\change}[2][1=]{\todo[linecolor=blue,backgroundcolor=blue!25,bordercolor=blue,#1]{#2}}
\newcommandx{\info}[2][1=]{\todo[linecolor=green,backgroundcolor=green!25,bordercolor=green,#1]{#2}}
\newcommandx{\improvement}[2][1=]{\todo[linecolor=orange,backgroundcolor=orange!25,bordercolor=orange,#1]{#2}}

\begin{document}
	
		\title[Denoising GW signals from BBHs with dilated convolutional AE]{Denoising gravitational-wave signals from binary black holes with dilated convolutional autoencoder}
		
		\author{Philippe Bacon}
		\address{Universit$\acute{e}$ Paris Cité, CNRS, Astroparticule et Cosmologie, F-75013 Paris, France}

		\author{Agata Trovato}
		\address{Dipartimento di Fisica, Università di Trieste, I-34127, Trieste, Italy; INFN, Sezione di Trieste, I-34127 Trieste, Italy}

		\author{Micha{\l} Bejger}
		\address{INFN Sezione di Ferrara, Via Saragat 1, 44122 Ferrara, Italy; Nicolaus Copernicus Astronomical Center, Polish Academy of Sciences, Bartycka 18, 00-716 Warszawa, Poland}
		
		\begin{abstract}
			Broadband frequency output of gravitational-wave detectors is a non-stationary and non-Gaussian time series data stream dominated by noise populated by local disturbances and transient artifacts, which evolve on the same timescale as the gravitational-wave signals and may corrupt the astrophysical information. We study a denoising algorithm dedicated to expose the astrophysical signals by employing a convolutional neural network in the encoder-decoder configuration, i.e. apply the denoising procedure of coalescing binary black hole signals in the publicly available LIGO O1 time series strain data. The denoising convolutional autoencoder neural network is trained on a dataset of simulated astrophysical signals injected into the real detector's noise and a dataset of detector noise artifacts (''glitches''), and its fidelity is tested on real gravitational-wave events from O1 and O2 LIGO-Virgo observing runs. 
		\end{abstract}
		
\vspace{1pc}

	
	
\section{Introduction}
\label{sect:intro}
	
The onset of gravitational wave (GW) astronomy began in 2015 with first direct detection of GWs from a binary inspiral and merger of two stellar-mass black holes (BHs), the event later denoted as GW150914 \cite{PhysRevLett.116.061102}. Since then, the Advanced LIGO \cite{2015CQGra..32g4001L} and the Advanced Virgo \cite{2015CQGra..32b4001A} detector network has detected many GW signals, mainly binary black holes but also binary neutron star (NS) inspiral events  \cite{2017PhRvL.119p1101A,2020arXiv200101761T}, and BH-NS systems \cite{Abbott_2020,2021ApJ...915L...5A}; for the list of published GW transient signal detections, see the GWTC-1 \cite{2019PhRvX...9c1040A}, GWTC-2 \cite{2021PhRvX..11b1053A}, GWTC-2.1 \cite{2021arXiv210801045T} and GWTC-3 \cite{2021arXiv211103606T} catalogs. During the latest observational campaign (LIGO-Virgo O3 run, 1st April 2019 - 26 March 2020), the LIGO-Virgo network was reporting GW signals with a rate of approximately once per week, with the alerts on highly-significant events (sky localization, type of source) released to the scientific community in the low-latency mode \cite{gracedb}. The detection rate will steadily increase with the increasing instrumental sensitivity, approaching one binary BH detection per a few days in the incoming O4 run \cite{2018LRR....21....3A}. Therefore, a rapid and reliable selection algorithm of data periods in which signals are hidden will be very useful standalone, or as a part of a broader detection or data characterisation framework. 

Raw GW strain data are fundamentally noisy time series in which the GW signals are hidden; for the detailed description of the GW data, see \cite{2019arXiv190811170T,2021SoftX..1300658A}. Classically, in order to confirm the existence of a signal in the data time-series, one has to apply a matched filtering technique (e.g., \cite{wienerbook,PhysRevD.44.3819,1999PhRvD..60b2002O}), which is an optimal technique only when the background noise is Gaussian and stationary, and which requires substantial computational resources, and a fine grid of filter templates build from GW signals parameters, to find the match. 

In this work we are studying an alternative to the established data-processing methods, and employ the machine learning (ML) approach to the GW signal detection and signal processing. Several applications of ML in the GW astronomy low-latency data analysis exist; see e.g., \cite{Huerta:2019rtg,PhysRevLett.120.141103,2018PhLB..778...64G} for specific applications in the context of binary BH searches, as well as \cite{10.1088/2632-2153/abb93a} for a recent review of ML in GW domain. ML methods are uniquely suited to identify patterns in data, but also to perform other data processing tasks, such as the {\em denoising}. Denoising of the GW data was applied in the past using the total-variation method \cite{2014PhRvD..90h4029T,2018PhRvD..98h4013T}, with the split Bergman regularization to obtain the total-variation regularization \cite{2015ASSP...40..289T}, and with the dictionary learning \cite{2016PhRvD..94l4040T}. From the deep neural network (NN) point-of-view, denoising methods were applied in \cite{2020PhLB..80035081W} with the {\tt WaveNet} implementation \cite{2016arXiv160903499V}, as well as using the auto-encoder (AE) architecture \cite{2018APS..APRS14008S,2019arXiv190303105S} to perform the denoising task, i.e. as a denoising auto-encoder (DAE) \cite{HintonSalakhutdinov2006b}, where instead of encoding and subsequently decoding an input sequence to itself, the training consists of feeding the noisy (``corrupted'') input and expecting a noiseless (``clean'') output. Recent works that apply this paradigm include \cite{2017arXiv171109919S} with the Long Short-Term Memory/ recurrent neural networks (LSTM/RNN, see e.g. \cite{10.5555/553011,2013arXiv1312.6026P,10.1162/neco.1997.9.8.1735}) concept, and \cite{2021PhRvD.104f4046C}, using a combination of the convolutional NN (CNN, see e.g. \cite{GU2018354,Goodfellow2016}) and LSTM. An algorithm implemented in \cite{2020Senso..20.6920L}, based on the local polynomial approximation combined with the relative intersection of confidence intervals rule for the filter support selection is applied to denoise the GW burst signals emitted during core collapse supernov{\ae} events. 

Here we implement a purposefully simple version of the DAE, based on one-dimensional input CNN paradigm, and apply it to the noisy (``corrupted'') time series GW data containing astrophysical signals immersed in the real noise, in order to study limitations in recovering the noiseless (``clean'') GW signals in this realistic setup. The CNN-DAE approach has advantages over implementations of DAE already existing in the literature, the primary being the fact that the CNN implementation is smaller and trains faster than recurrent NN. We demonstrate that a relatively small CNN DAE with a few dilated decoder layers \cite{yu2016multiscale} is able to train on the GW signal waveforms injected to realistic LIGO data time series, and recover real GW events. We consider this type of a lightweight algorithm a potentially useful trigger generator, performing a role of rapid initial classification of GW signals, and/or data characterisation tasks.

This article is composed as follows. Section~\ref{sect:methods} contains a brief description of the CNN and AE methodology, description of the DAE network, as well as the training and testing data. Section~\ref{sect:results} describes the results, obtained using both the simulated signals and real O1 and O2 events, whereas Sect.~\ref{sect:conclusions} contains the conclusions. 

\section{Methods and training data}
\label{sect:methods}

\subsection{Convolutional autoencoders}

CNN is a type of NNs that applies a set of convolutions (by means of the kernel filters) to the network's input \cite{GU2018354,Goodfellow2016}. In the context of denoising, training a CNN consists in finding the filters weights which optimally extract the preferable features in input data and are present in clean (noise-free) data. The filters are weighted, and their relative importance is also acquired during training. CNNs are widely used in data processing and are able to perform a wide range of tasks such as classification, pattern recognition or feature extraction. Concretely, the convolutional filter is moved across the input image to reveal important features while preserving spatial relationship between samples, see e.g. \cite{Dhillon2019ConvolutionalNN,Yao2019ARO}. In all cases, the CNN is effectively performing a role of a dimensionality reduction algorithm, the output of the convolutional layer being a `simplified view' of the input. In our one-dimensional context of time series data we adopt the CNN architecture for a practical advantage of their much faster training times, as compared to more complicated LSTM/RNN architectures \cite{Goodfellow2016}, while retaining the ability to learn temporal information encoded in the time series \cite{2018arXiv180904356I}. Specifically, we introduce dilated convolutional layers \cite{yu2016multiscale}, with an aim to connect distant samples within the signal, and use multi-scale information contained in the time series.

\subsection{Denoising AE model}
\label{sect:denoising_ae} 

The CNN layers in this work adopt the AE architecture. In general, the AE NN represents an identity function by compressing the representation of input data and then decompressing it. During the training, the overall network will learn its own sparse representation of the input signals. AE are composed of two networks: an encoder $g_{\phi}$, and a decoder $f_{\theta}$.

\tikzset{arrow/.style={-stealth, thick, draw=gray!80!black}}
\tikzset{cross/.style={cross out, draw=black, minimum size=2*(#1-\pgflinewidth), inner sep=0pt, outer sep=0pt}, cross/.default={4pt}}
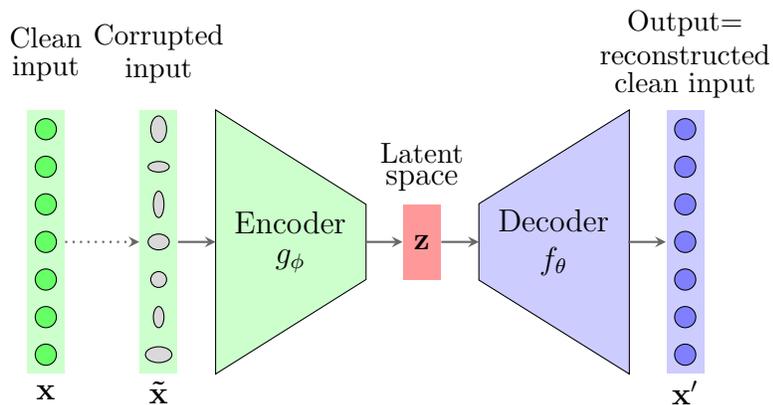
\begin{figure}[ht]
\centering
\begin{tikzpicture}
       
  \node[fill=green!20, minimum width=0.5cm, minimum height=3.5cm] (Xo) at (-1.5,0) {};
  \draw[fill=green!60] (-1.5,-1.5) circle (4pt);
  \draw[fill=green!60] (-1.5,-1) circle (4pt);
  \draw[fill=green!60] (-1.5,-0.5) circle (4pt);
  \draw[fill=green!60] (-1.5,0) circle (4pt);
  \draw[fill=green!60] (-1.5,0.5) circle (4pt);
  \draw[fill=green!60] (-1.5,1) circle (4pt);
  \draw[fill=green!60] (-1.5,1.5) circle (4pt);

  \node[fill=green!20, minimum width=0.5cm, minimum height=3.5cm] (X) at (0,0) {};

  \node at (-1.5,-2) {$\mathbf{x}$}; 

  \node at (0,-2) {$\mathbf{\tilde{x}}$}; 

  \draw[fill=gray!30] (0,-1.5) ellipse (5pt and 3pt);
  \draw[fill=gray!30] (0,-1) ellipse (2pt and 4pt);
  \draw[fill=gray!30] (0,-0.5) ellipse (3pt and 3pt);
  \draw[fill=gray!30] (0,0) ellipse (4pt and 3pt);
  \draw[fill=gray!30] (0,0.5) ellipse (2pt and 5pt);
  \draw[fill=gray!30] (0,1) ellipse (4pt and 2pt);
  \draw[fill=gray!30] (0,1.5) ellipse (3pt and 5pt);

  \draw[fill=green!20] ([xshift=0.5cm]X.north east) -- ([xshift=2.5cm,yshift=0.5cm]X.east) -- ([xshift=2.5cm,yshift=-0.5cm]X.east) -- ([xshift=0.5cm]X.south east) -- cycle; 

  \node at (1.75,0) {\shortstack{Encoder\\~\\$g_\phi$}}; 
 
  \node[fill=red!40, minimum width=0.5cm, minimum height=1.0cm] (Z) at (3.5cm,0) {$\mathbf z$};
  
  \draw[fill=blue!20] ([xshift=0.5cm]Z.north east) -- ([xshift=2.5cm,yshift=1.25cm]Z.north east) -- ([xshift=2.5cm,yshift=-1.25cm]Z.south east) -- ([xshift=0.5cm]Z.south east) -- cycle;

  \node at (5.25,0) {\shortstack{Decoder\\~\\$f_\theta$}};
  
  \node[fill=blue!20, minimum width=0.5cm, minimum height=3.5cm] (Xp) at (7,0) {};
  \node at (7,-2) {$\mathbf{x'}$};

  \node[above of=Z, minimum width=0.5cm, minimum height=2.0cm] (B) {\shortstack{\small Latent\\ space}};  

  \node[minimum width=0.5cm, minimum height=1.0cm] (I) at (-1.5, 2.5) {\small\shortstack{Clean\\input}};  
  \node[minimum width=0.5cm, minimum height=1.0cm] (I) at (0, 2.5) {\small\shortstack{Corrupted\\input}};  

  \node[minimum width=0.5cm, minimum height=1.0cm] (O) at (7,2.5) {\small\shortstack{Output=\\reconstructed\\clean input}};  
  \draw[fill=blue!50] (7,-1.5) circle (4pt);
  \draw[fill=blue!50] (7,-1) circle (4pt);
  \draw[fill=blue!50] (7,-0.5) circle (4pt);
  \draw[fill=blue!50] (7,0) circle (4pt);
  \draw[fill=blue!50] (7,0.5) circle (4pt);
  \draw[fill=blue!50] (7,1) circle (4pt);
  \draw[fill=blue!50] (7,1.5) circle (4pt);
 
  \draw[arrow] (X.east) -- ([xshift=0.5cm]X.east);
  \draw[arrow] ([xshift=-0.5cm]Z.west) -- (Z.west);
  \draw[arrow] (Z.east) -- ([xshift=0.5cm]Z.east);
  \draw[arrow] ([xshift=-0.5cm]Xp.west) -- (Xp.west);
  \draw[arrow,dotted] (Xo.east) -- (X.west);
    
\end{tikzpicture}

\caption{Data flow diagram of a DAE architecture. The clean input data $\mathbf{x}$ is corrupted (denoted by $\mathbf{\tilde{x}}$) and fed to the encoder part of the AE; for the well-trained network, the decoded output (reconstructed clean input) $\mathbf{x'}$ should be a close analogue of the clean input $\mathbf{x}$, i.e. $\mathbf{x'}\approx \mathbf{x}$. The middle part of the DAE - the latent space - is denoted by $\mathbf{z}$. In case of $\mathbf{x}\equiv\mathbf{\tilde{x}}$ the DAE becomes a classical AE.} 
\label{fig:dae}
\end{figure} 

The parameters $\phi$ and $\theta$ denote the parameters of the encoder and the decoder respectively. The encoder maps the original high-dimensional input into a {\em latent space} $\mathbf{z} = g_{\phi}(\mathbf{x})$, where $\mathbf{x}$ is the training or testing input. Note that we do not make an explicit use of the latent space in the present study, although in principle it may be used for e.g. parameter estimation study, as it is related to the Bayesian formulation of the problem through the variational AE (VAE) approach \cite{2013arXiv1312.6114K,Kingma+2019}, or a version of the VAE, conditioned by actual values of signal parameters in question (Conditional VAE, CVAE \cite{NIPS2015_8d55a249}). Contrary to the encoder, the decoder recovers the data from $\mathbf{z}$, and successively unfold the compressed data. The output data of the decoder (and hence the AE) is denoted $\mathbf{x'} = f_{\theta}(\mathbf{z})$. AE accomplishes the dimensionality reduction in the same way as the principal component analysis (PCA, \cite{Jolliffe2011}) or matrix factorisation algorithms \cite{10.5555/3008751.3008829}, but the underlying process of AEs is highly non-linear and explicitly optimizes the data reconstruction.

The training of an AE consists in finding the set of parameters $(\theta, \phi)$ which minimize the distance between $\mathbf{x}$ and $\mathbf{x'}$. This distance is properly defined by the so-called {\it loss function} $L_{AE}(\theta, \phi)$. In the case of regression problems like the AE identity function, a popular choice is the mean square error (MSE):  
\begin{equation}
	L_{AE}(\theta, \phi) = \sum_{i=1}^{N} \left( x_{i} - x'_i \right)^2 = \sum_{i=1}^{N} \left( x_{i} - f_{\theta}(g_{\phi}(x_{i})) \right)^2, 
  \label{eq:ae_loss_func}
\end{equation}
where $N$ is the number of samples of input/output signals.

We adopt a one-dimensional CNN AE to denoise generated astrophysical GW signals added to real GW detector data time series, collected during the LIGO O1 run. A DAE procedure consists in training an AE with a corrupted version of the input signal (i.e. clean signal immersed in the noisy time series), denoted by $\mathbf{\tilde{x}}$, and by demanding that the recovered output $\mathbf{x'}$ is as close to the original clean input $\mathbf{x}$ as possible. A schematic representation of a DAE is presented in Fig.~\ref{fig:dae}. The training data set is described in detail in the subsequent Sect.~\ref{sect:training_data}. Results of the trained DAE on simulated data in real detector's noise are described in Sect.~\ref{sect:results}, whereas denoising of {\em real} GW signals found in the O1 and O2 LIGO run are presented in Sect.~\ref{sect:real_gw}.

\begin{table}[ht]
    \centering    
    \renewcommand{\arraystretch}{0.7}
    {
    \begin{tabular}{c c c c}
        Layer number & Layer type & Output shape \\
        \cmidrule{1-3}
        1 & Input & ($N$) & \rdelim\}{8}{*}[{\rotatebox[origin=c]{90}{encoder}}] \\
        2 & BatchNormalisation & ($N$) \\
        3 & Reshape & ($N$, 1)  \\
        4 & Conv1D (128 units) & ($N$, 128) \\
        5 & MaxPooling1D & ($N/2$, 128) \\
        6 & Conv1D (64 units) & ($N/2$, 64)  \\
        7 & MaxPooling1D & ($N/4$, 64) \\
        8 & Conv1D (32 units) & ($N/4$, 32) \\
        \cmidrule{1-3}
        9 & Conv1D (32 units) & ($N/4$, 32) & \rdelim\}{11}{*}[{\rotatebox[origin=c]{90}{decoder}}] \\
        10 & UpSampling1D & ($N/2$, 32) \\
        11 & Conv1D (64 units) & ($N/2$, 64) \\
        12 & UpSampling1D & ($N$, 64) \\
        13 & Conv1D (128 units) & ($N$, 128) \\ 
        \cmidrule{1-3}
        14 & Conv1D (128 units, DR=2) & ($N$, 128) & \hspace{5em}\rdelim\}{4}{*}[{\rotatebox[origin=c]{90}{dilation}}] \\
        15 & Conv1D (128 units, DR=3) & ($N$, 128) \\
        16 & Conv1D (128 units, DR=4) & ($N$, 128) \\
        17 & Conv1D (128 units, DR=8) & ($N$, 128) \\
        18 & Dense (1 unit) & ($N$, 1) \\
        19 & Reshape & ($N$) \\
    \end{tabular}
    }
    \renewcommand{\arraystretch}{1}
\caption{Architecture of the DAE model. Here $N=2048$ denotes the number of samples in the time series signal instance provided at the input of the encoder (the number of samples corresponds to 1 second of data with the frequency sampling of $2048$ Hz). The AE latent space ''bottleneck'' is located between layers 8 and 9. DR stands for dilation rate. The dilated part of the decoder starts at layer 14.}
    \label{tab:dae_archi}
\end{table}

The layer-by-layer structure of the NN is described in detail in Tab.~\ref{tab:dae_archi}. The batch normalization technique \cite{ioffe2015batch} normalizes and scales the layer inputs in order to stabilize (prevent saturation of the non-linearities), and to speed up the training. The overall effect is also to make the network more robust to the initialization of weights. We use 32 input instances per batch. In the encoder part of the AE, the pooling layers \cite{maxpool2010} apply a non-linear down-sampling and compactify (reduce) the information of the input. In the decoder part of the AE, we introduce upsampling layers to ensure that the low dimensional information is successively unfolded. We use a Rectified Linear Unit (ReLU) activation function all over the layers. As there are no strong arguments for the use of asymmetric encoder-decoder structure, we introduce three dilated convolutional layers \cite{yu2016multiscale} which correlate non adjacent samples within the signal, and is able to systematically gather multi-scale information contained in the time series without losing resolution. In this respect the final layers (CNN dilated layers) of the decoder bear resemblance to the LSTM architecture, in the sense that they register and keep track of the signal evolution in various scales. More precisely, they include a causal padding that couples signal samples that are originally far from each other. The more important is the dilation rate, the wider is the coupling area. 

For the DAE loss function, we chose the following MSE function:
\begin{equation} 
    L_{DAE}(\theta, \phi) = \sum_{i=1}^{N} \left( x_{i} - f_{\theta}(g_{\phi}(\tilde{x}_{i})) \right)^2, 
    \label{eq:dae_loss_func} 
\end{equation}
where $f_{\theta}(g_{\phi}(\mathbf{\tilde{x}}))=\mathbf{x'}$ is the DAE output, and $\mathbf{x}$ is the ground-truth (clean) signal waveform input.

\subsection{Training and testing data}
\label{sect:training_data}
The input ``clean'' signals used in this project are simulated astrophysical GW signals from binary BHs (BBHs). In general, astrophysical GW signals from close binary compact systems exhibit a characteristic increase of GW amplitude and frequency during the inspiral (the ``chirp''), followed by the merger of the binary components, and the ringdown GW emission from the remnant \cite{2017AnP...52900209A}. Approximately, the GW inspiral frequency evolves as \cite{GW150914-basic2016}
\begin{eqnarray}
f_{GW}^{-8/3}(t) = \frac{(8\pi)^{8/3}}{5}\left(\frac{G\mathcal{M}_c}{c^3}\right)^{5/3}\left(t_c-t\right)\ \textrm{+ higher order corrections},
    \label{eq:chirp}
\end{eqnarray}
where $t_c$ is the time of coalescence, and the $\mathcal{M}_c$ is a function of component masses $M_1$ and $M_2$, called the \textit{chirp mass}:
\begin{equation} 
    \mathcal{M}_c = \frac{(M_1 M_2)^{3/5}}{(M_1 + M_2)^{1/5}}. 
    \label{eq:chirp_mass} 
\end{equation}
At a given moment the GW strain amplitude $h$ depends, approximately, on the binary system parameters as follows: 
\begin{equation} 
h(r) \propto \mathcal{M}_c^{5/3} f_{GW}^{2/3} / r. 
    \label{eq:gw_amplitude}
\end{equation}
For production runs, we assume that signal waveforms (the amplitude-frequency evolution $h(f)$ or, equivalently, $h(t)$) are well-modeled using general relativity, and for the sake of this study employ the SEOBNRv4 waveform model \cite{2017PhRvD..95d4028B}, assuming non-spinning components with masses randomly chosen from a uniform distribution in the range $M_1, M_2 \in (10,\ 30)\,M_\odot$, compatible with the current state of observational knowledge on the binary BH population \cite{2019PhRvX...9c1040A}. Sky localizations were chosen to be optimal with respect to the antenna pattern of the detector for a given time, corresponding to the data segment. Other parameters describing the source, e.g., the orientation of the orbit with respect to the line of sight, were selected randomly from their natural ranges.

The ``corrupted'' input is obtained injecting these astrophysical GW signals in pre-selected time series segments from the science-quality data stream of the LIGO Livingston detector, collected during the O1 data taking campaign and released by the LIGO and Virgo Collaborations via the Gravitational Wave Open Science Center \cite{2021SoftX..1300658A}.
For convenience, these time series segments are packaged with a fixed length of 1 s and down-sampled from the detector's raw sampling rate of 16384 Hz to 2048 Hz (corresponding to the Nyquist frequency of 1024 Hz). They are selected to contain only ``quiet'' data, i.e. rejecting periods when transient artifacts of instrumental origins (so-called ``glitches'') \cite{transLIGO,2017CQGra..34f4003Z}, as well as hardware signal injections (\cite{inj}, arranged to test and calibrate the detection system), are present. The GW signals are injected in the time series segments randomly off-center (taking the signal's maximum amplitude peak as a reference), with random time shifts drawn from a uniform distribution of (-0.1, 0.4) s.

GW data analysis is essentially based on matched-filtering techniques, which consist in finding the waveform template that matches the data best. Working under the assumption of the additive property of the noise (i.e., the time series $d$ containing the GW signal $h$ immersed in noise $n$ is $d = n + h$), then the optimal signal-to-noise ratio (SNR) $\rho$, corresponding to the best matching filter (template $h$ equals the signal) is 
\begin{equation}     
    \rho = \frac{(d|h)}{\sqrt{(h|h)}},\quad\mathrm{where}\quad (d|h) = 4\Re\int\frac{\tilde{d}(f)\tilde{h}^*(f)}{S_n(f)}\mathrm{d}f,
\label{eq:snr}    
\end{equation} 
with $\tilde{d}(f)$ a Fourier transform of the time series $d(t)$, and $S_n(f)$ the power spectral density (PSD) of the detector; asterisk denotes complex conjugation \cite{1999PhRvD..60b2002O}. Detector's PSD $S_n(f)$ represents the frequency-dependent sensitivity in a broad range of frequencies, and is quantified by its sensitivity curve. 
From the astrophysical perspective, the SNR is a function of the waveform amplitude and, since the waveform describes the evolution of the GW amplitude, is inversely proportional to the luminosity (''loudness'') distance. While preparing the data set, we label the signal waveform with their {\it optimal} matched-filter SNR 
\begin{equation}
\rho_{opt}=\sqrt{(h|h)} = \sqrt{ 4\int\nolimits_0^\infty {{{\vert \tilde h(f){\vert ^2}} \over {{S_n}(f)}}{\rm{d}}f}}, 
\label{eq:snr_opt}
\end{equation} 
which approximates $\rho$, assuming that the noise effect is negligible, $d\approx h$. The optimal matched-filter SNR $\rho_{opt}$ is a good first order approximation to the actual matched-filter SNR $\rho$ in e.g. stationary Gaussian noise \cite{2012LRR....15....4J}. Subsequently, we produce a synthetic distribution of source luminosity distances such that the $\rho_{opt}$ distribution is uniform in the range of $(5,\ 20)$ in order to consider a wide range of signals during the training. Source distances are in the range between 100 to 1000 Mpc. The adopted $\rho_{opt}$ range is comparable to the figure-of-merit optimal SNR of 8, which correspond to a confident single-interferometer detection of an optimally-oriented compact binary inspiral \cite{PhysRevD.47.2198,2010CQGra..27q3001A}, and is also consistent with the previous LIGO-Virgo detections \cite{2019PhRvX...9c1040A,2021PhRvX..11b1053A,2021arXiv210801045T}. We extend the $\rho_{opt}$ range to the lower values to study the sensitivity and robustness of our DAE implementation in the low $\rho$ regime. 

\begin{figure}[ht] 
    \centering
    \includegraphics[width=\columnwidth,clip]{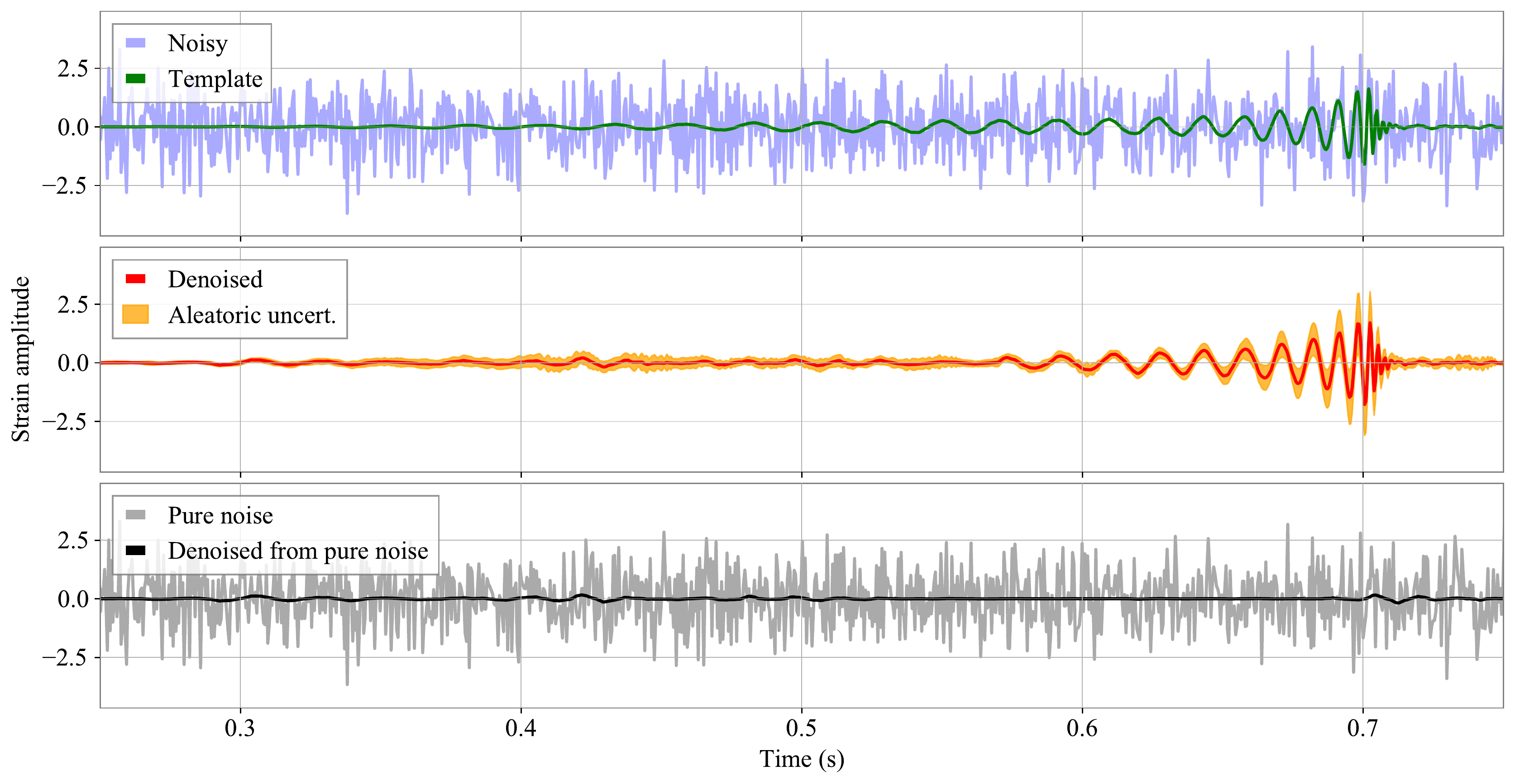}
    \caption{An example of a training/testing data time series instance, and the DAE output. Top panel: a GW signal (green) immersed in the noisy data (blue). Middle panel: denoised version of the signal (red), with the aleatoric uncertainty estimate (yellow); see Sect.~\ref{sect:aleatoric_uncertainty} for more details. Bottom panel: for tests, we perform a denoising procedure on the time series without a GW signal added (grey). The result is the black line, optimally with signal-to-noise equal to 0. Parameters of the GW waveform are as follows: $M_1=22.74\ M_\odot$, $M_2=17.34\ M_\odot$, distance $r=479.27$ Mpc, corresponding to the optimal matched filter signal-to-noise ratio $\rho_{opt}=9.3$ (Eq.~\ref{eq:snr_opt}). The denoised $\rho_{opt,d}$ is 10.3. Output waveform obtained from pure-noise sample (containing no GW signal) has $\rho_{opt,pn}$ of 1.5. Both the data and GW waveform are whitened by the detector's PSD, calculated using neighbouring data segments; see Sect.~\ref{sect:results} for more details.}
    \label{fig:waveform_example}
\end{figure}

 Last, but not least, to normalize the dependence of signals' strength at different frequencies we additionally perform the {\em whitening} of the time series data with added signals: we divide the Fourier representation of the time domain data by an estimate of the amplitude spectral density of the noise $\sqrt{S_n(f)}$ (ASD, square root of the PSD) to ensure that the data has equal significance in each frequency bin \cite{10.1088/1361-6382/ab685e}.  A low-frequency cutoff at $f_{low}=30$ Hz was chosen for the simulated signals and the data, taking into account the low frequency (seismic) limit of the detectors' sensitivity. This is reflected in the example of GW strain amplitude evolution $h(t)$, immersed in the LIGO Livingston detector noise, shown on Fig.~\ref{fig:waveform_example}.

The network is trained during 50 epochs on 7000 data time series containing injected astrophysical signals. In addition, the training set contains 1000 time series from the O1 LIGO Livingston data when known instrumental artifacts (''glitches'') are present in the data. We have used the the {\tt Gravity Spy} database \cite{2017CQGra..34f4003Z} to obtain various common types of glitches with an estimated SNR larger than 10. All the instances in this dataset are treated with the whitening procedure and a 30 Hz high band pass filter. Mixing signals and glitches prevents from any over-fitting because the network is fed with several distinct datasets which all span the targeted parameter space. Additionally, we introduce randomization of the data at two levels. First, the set of training and testing datasets are split in two after randomly shuffling the initial dataset; the train-test split factor is 0.75. In addition, we randomly replace some of the training and testing astrophysical signal input data by a null signal (array of zeros) with a probability $p = 35 \%$, to increase the robustness of the trained CNN to Gaussian noise fluctuations (in case of instrumental glitches, the instances of signal are automatically set to arrays containing zero values). In training we have used the Adam optimizer \cite{2014arXiv1412.6980K} with a constant learning rate of 0.001 as an optimization algorithm for stochastic gradient descent calculations.

\section{Results}
\label{sect:results}
Using the figures of merit commonly used in the GW astronomy, the MSE (Eq.~\ref{eq:dae_loss_func}), and the waveform overlap $\mathcal{O}$, which compares the original ''clean'' waveform $h$ and the denoised waveform  $h^d$ obtained at the output of the DAE: 
\begin{equation} 
    \mathcal{O} = \sqrt{{\sum^N_{i=0} h_i h^d_i}\left(\sum^N_{i=0} h_i h_i\right)^{-1}}, 
    \label{eq:overlap}
\end{equation} 
where $N$ is the number of points in the time series. In the evaluation, we also show distributions of selected astrophysical parameters of the GW signals. The distribution of the SNR of denoised signals is compared to the SNR of injected signals. As a sanity check we evaluate the DAE on time series not containing injected signals (''only noise'' time series) and on several types of the {\tt Gravity Spy} database \cite{2017CQGra..34f4003Z} glitches.   

\subsection{Population of injected signals}
\label{sect:injected_signals}

\begin{figure}[!ht]
        \begin{center}
        \includegraphics[scale=0.85]{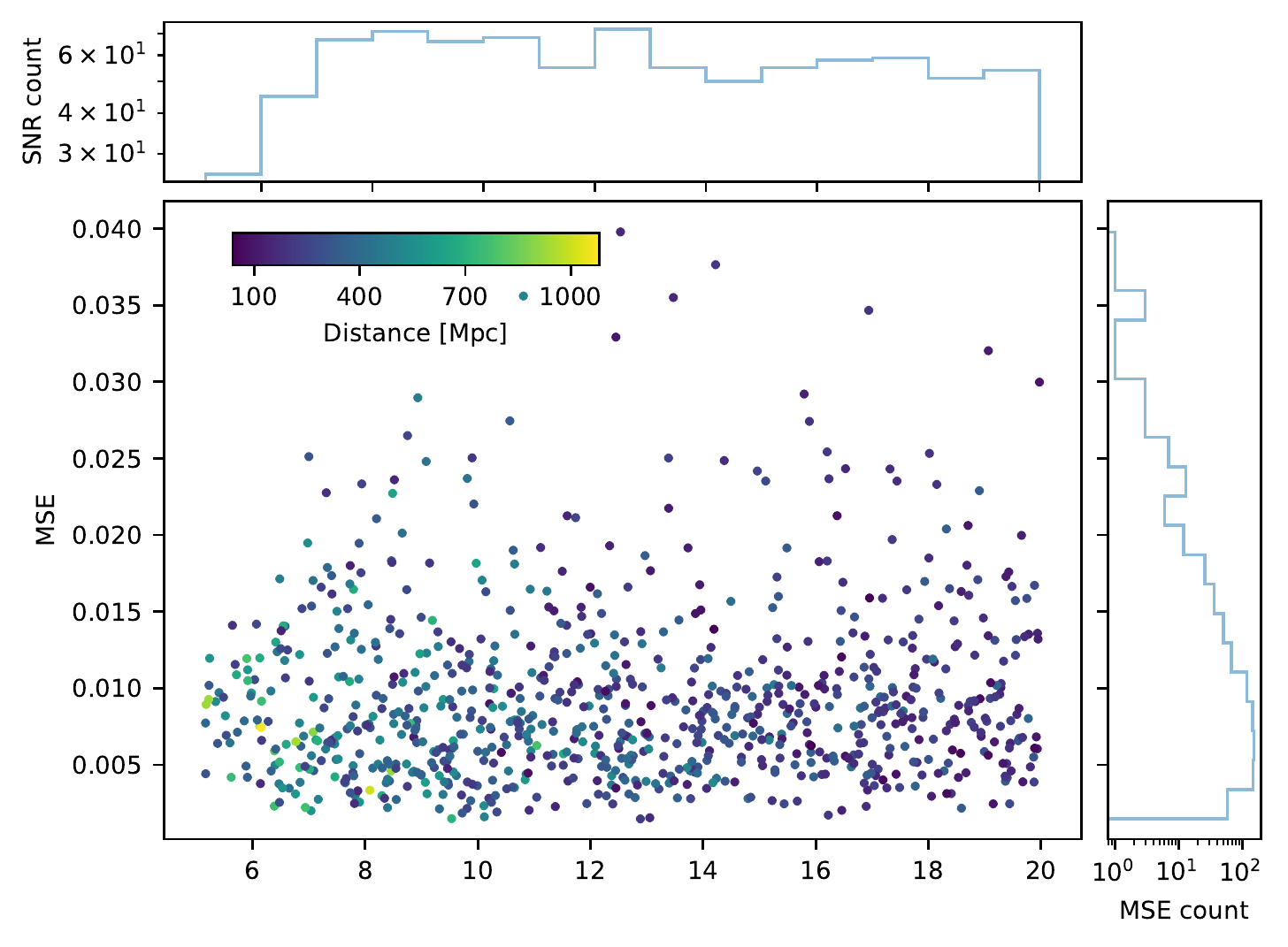}
		\vskip -5pt 
		\hskip 10pt
		\includegraphics[scale=0.85]{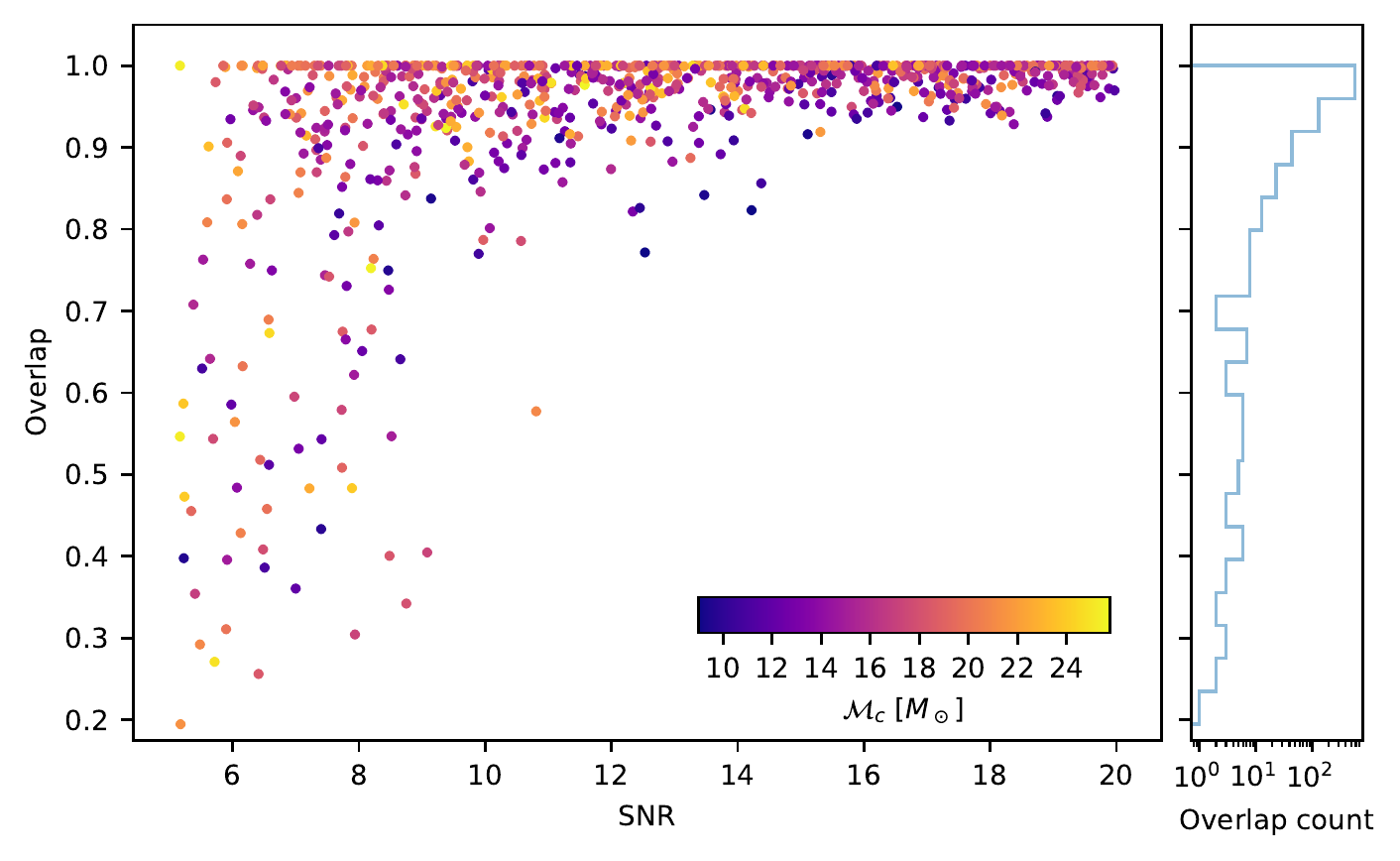}
		\end{center} 
		\caption{Upper part of the plot: mean square error (MSE, Eq.~\ref{eq:dae_loss_func}) as a function of the signal-to-noise (SNR) resulting from for evaluating 1000 data instances with added astrophysical GW waveforms. Color of points correspond to the source distance (in Mpc). Side panels show histogram counts (in logarithmic scale) along both axes. Lower part of the plot: the overlap (Eq.~\ref{eq:overlap}) between the original and the denoised GW waveforms as a function of the SNR. Color of points correspond to the chirp mass $\mathcal{M}_c$. The histogram count of the SNR values shows approximately uniform distribution of the values in the dataset.}
		\label{fig:mse_ove_vs_snr}
\end{figure}

In Fig.~\ref{fig:mse_ove_vs_snr} we show both the estimated MSE and the recovered waveform overlap $\mathcal{O}$ as a function of the injected matched filtering signal-to-noise ratio (SNR) \cite{1999PhRvD..60b2002O, 2020PhLB..80035081W}, with additional information on the distance to the source (top plot) and the chirp mass (bottom panel) encoded in color, for 1000 GW signals added to detector's noise. We don't detect a correlation between the MSE and the SNR, as expected in properly executed training on the input data set with uniform distribution SNR. While we don't detect a clear correlation between the overlap and the chirp mass $\mathcal{M}_c$, a preference towards smaller SNR for large distances is seen. Signals with waveform overlap smaller than 0.75 constitute 6.6\% of all the signals; 5.6\% of all the signals have both $\mathcal{O} < 0.8$ and SNR $<8$, so using this threshold overlap criterion we estimate that about 1\% of potentially detectable signals (with SNR $>8$) are incorrectly recovered by the DAE. 

\begin{figure}[!ht]
        \begin{center}
        \includegraphics[scale=0.85]{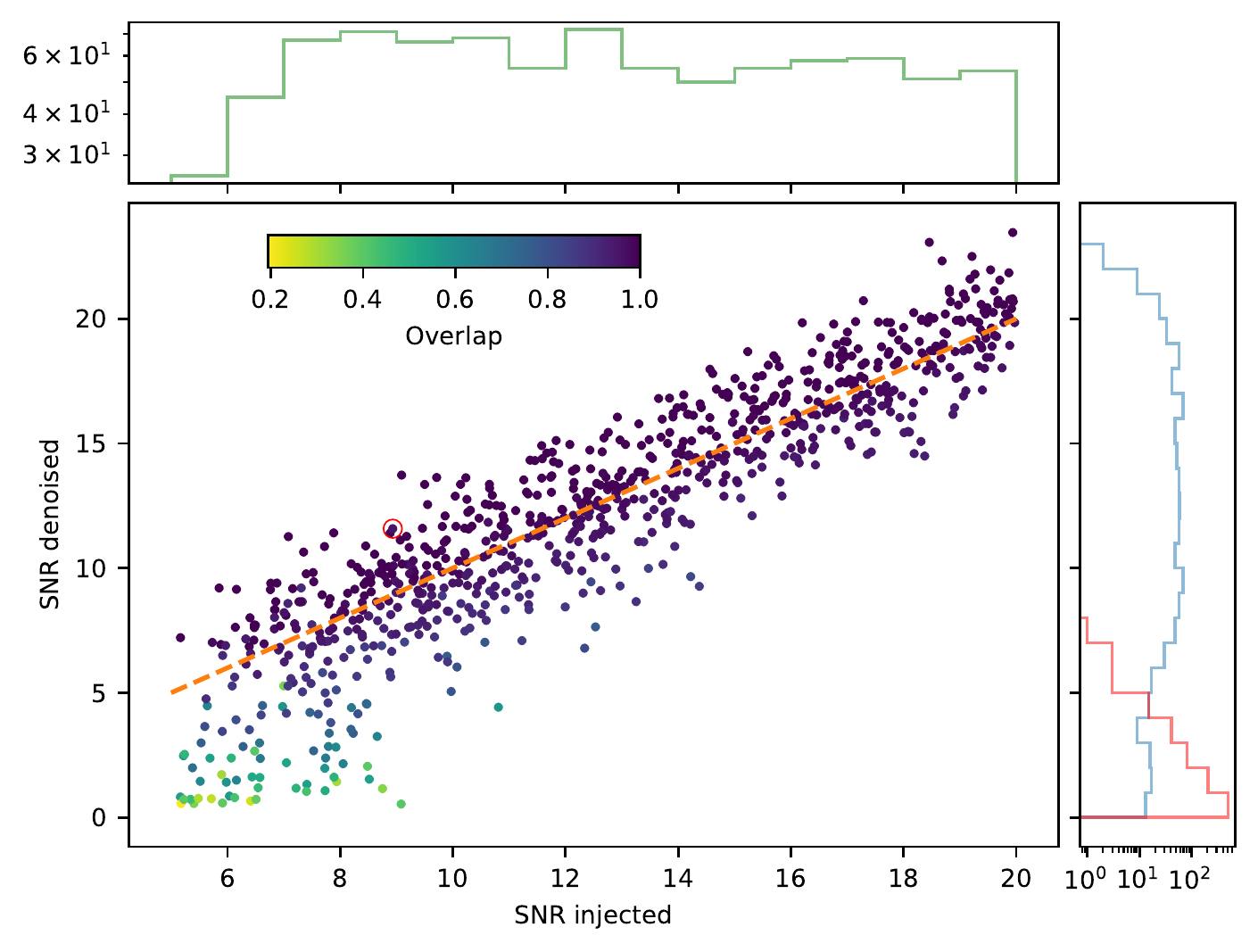}
		\end{center} 
		\caption{Denoised SNR (calculated from the DAE output, vertical axis) as a function of the injected SNR (horizontal axis) for a testing dataset of 1000 data instances with added astrophysical GW waveforms. Points are colored by their corresponding overlap values. Orange dashed line denotes the denoised SNR equal to injected SNR. Example waveform presented in Fig.~\ref{fig:waveform_example} is denoted by a red circle. Side histograms (in logarithmic scale) show the distribution of the injected SNR (upper plot), and SNRs denoised from samples containing added GW waveforms (blue histogram), and - for comparison - not containing GW signals (i.e. pure noise, red histogram), respectively.} 
		\label{fig:snri_snrd_snrdno}
\end{figure}

Figure~\ref{fig:snri_snrd_snrdno} shows the comparison between the injected SNR $\rho_{opt}$ and the recovered (denoised) output SNR $\rho_{out,d}$, both calculated using the optimal matched filter SNR  formula (Eq. \ref{eq:snr_opt}). The color code indicates the waveforms overlap. As expected, in cases of high overlap the denoised SNR approximates quite well the injected one. The cases of lower overlap have a denoised SNR close to zero. The distribution follows the ideal $\rho_{out}\equiv \rho_{out,d}$ relation with a root-mean-square of residuals of 1.9 and variance of residuals of 3.8. The denoised SNR may be used as an approximate proxy for detection criterion: 8.2\% of the output  signals have $\rho_{opt,d}<5$, whereas 1.3\% of signals have both $\rho_{opt}>8$ and $\rho_{opt,d}<5$, i.e. are potentially strong enough to detect with the standard methods, but incorrectly recovered by the DAE. Additionally, we perform the denosing procedure on the same detector time series samples but {\em without} injected GW signals, to study the output signals. The distribution of output SNR in that case is depicted by the red histogram; only a few noise-only samples have $\rho_{out,d}>5$.

\subsection{Evaluation on known instrumental glitches}
\label{sect:glitches}

As discussed in Sect. \ref{sect:training_data}, we have also used time series containing known glitches to train the network.  
The GPS times of these glitches have been extracted from the {\tt Gravity Spy} database \cite{2017CQGra..34f4003Z}. The corresponding 1 second LIGO data segments centered at these GPS times have been downloaded via the Gravitational Wave Open Science Center \cite{2021SoftX..1300658A}.
For the training phase, 1000 GPS times have been randomly selected among the list of glitch times available in O1 with SNR$>$10 and duration$<$1 s.
To test the ability of the network to denoise also time series containing glitches, we have tested the network on different glitch families, selecting some examples of recurrent glitch types, still with the conditions SNR$>$10 and duration$<$1 s. The nomenclature of these families comes from {\tt Gravity Spy}. In some cases the origin of the glitch is known, for example for the glitch type called \textit{whistle}, which seems to be caused by signals at megahertz frequencies that beat with Voltage Controlled Oscillators in the interferometer control system \cite{whistle}. The other glitch families considered for this test, \textit{Low-Frequency Burst}, \textit{Koi Fish} and \textit{Blip}, have an unknown origin. A Low-Frequency Burst appears in a time-frequency spectrogram as an excess noise at low frequency. A Koi Fish is a short-duration broadband noise. A blip is a short duration noise that appears in a spectrogram as a symmetric ‘teardrop’ typically between 30 and 250 Hz, with the majority of the power appearing at the lowest frequencies \cite{transLIGO}.

\begin{figure}[!ht]
        \begin{center}
        \includegraphics[scale=0.85]{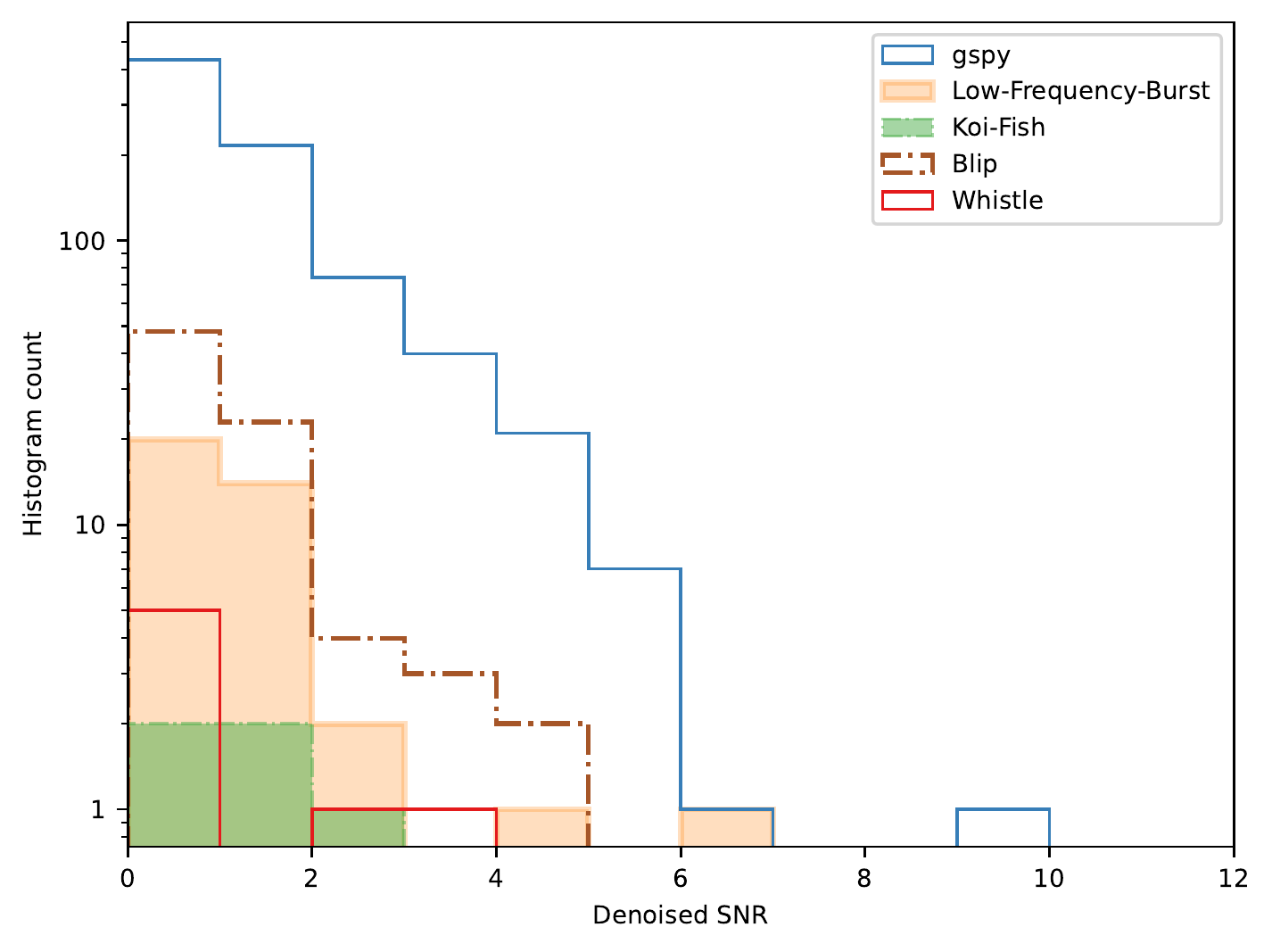}
		\end{center} 
		\caption{Evaluation of the DAE on instrumental glitches. The logarithmic vertical scale plot shows histograms of denoised output SNR for a selection of 38 Low Frequency Burst glitches, 5 Koi Fish type glitches, 80 Blips and 7 Whistle glitches. Blue line marks the evaluation on 792 assorted various types of glitches. All the glitches data are obtained from the {\tt Gravity Spy} database \cite{2017CQGra..34f4003Z}. The glitches have their estimated intrinsic SNR $>10$.}
	    \label{fig:glitches_snrd_hist}
\end{figure}

Figure~\ref{fig:glitches_snrd_hist} contains an histogram of the recovered SNR for a group of various glicthes randomly selected (blue line) and for specific glitch classes. It is visible that, even if all the selected glitches have SNR$>$10, the denoised SNR is always quite small, so it can be assumed that a denoised time series with sufficiently high SNR is probably a signal of cosmic orgin and not a glitch. However, this test have been done with few samples so a deeper study on this respect is needed. 

\subsection{Aleatoric uncertainty modelling}
\label{sect:aleatoric_uncertainty}

The results presented above are obtained with a forward pass of a noisy GW signal in our DAE model. It produces another denoised signal which has the same length as the input one. It is the result of minimizing an unweighted L2 distance between samples (Eq.~\ref{eq:dae_loss_func}) or equivalently of maximising a Gaussian likelihood. Obtaining a maximum likelihood estimator (MLE) provides an unbiased estimator of the mean. However this does not provide an unbiased estimator for the standard deviation.
\\When it comes to estimating uncertainty in a statistical model, two contributions are often distinguished. The \textit{epistemic uncertainty} is caused by the fact the model is not appropriate to the data. Indeed multiple model parameters could be consistent with the observed training data. In practice it arises in parts of the parameter space where there are fewer samples for training. The \textit{aleatoric uncertainty} is the uncertainty arising from the stochastic nature of observed data.

A further improvement of our work is notably to enrich the prediction made by the NN model by introducing the estimation of the aleatoric uncertainty.  To do so we follow the following procedure. First, the noisy signal $\tilde{x}(t)$ is fed into the DAE and it predicts a point estimate of the denoised signal $x'(t)$. Then we add a random noise $n(t)$ from a standard normal distribution to $x'(t)$ in order to form a new noisy signal.
Repeating the last step several time one gathers $\{ \tilde{x'}(t) \}_M$, i.e. $M$ distinct Gaussian noise realisations on top of $x'(t)$. Finally, the $M$ noisy timeseries are fed to the DAE and one gets $\{\tilde{x''}\}_M$ denoised timeseries from which the standard deviation at each time sample is computed.

This estimation of the aleatoric uncertainty relies on two assumptions. First, the noise part of the input signal follows a Gaussian distribution. Second, the point estimate is an unbiased estimate of the mean true signal. On Fig.~\ref{fig:waveform_example} and Figs.~\ref{fig:GW150914}-\ref{fig:GW170608_50Hz}, this mean is effectively close from the clean signal. No quantitative estimation of the closeness of the mean with the true signal has been performed.

Recent works on Bayesian deep learning like normalizing flows \cite{rezende2016variational} and variational autoencoders \cite{Kingma+2019} offer a formalized statistical and probabilistic framework to properly estimate both the epistemic and aleatoric uncertainties. This implies estimating a distribution on every sample of the signal. For instance, if we assume Gaussian distributed prediction of a denoising CVAE model, this implies the inference of $\left\{ \mu[s], \sigma[s] \right\}_{s}$ for $s$ ranging from $0$ to $2048$. We leave this investigation for further work.

\subsection{Real GW events}
\label{sect:real_gw} 

The result of the denoising of data segments containing real GW events registered during O1 and O2 are showed in \ref{sect:appendixO1} and \ref{sect:appendixO2}, respectively. We study events included in the {\tt GWTC-1-confident} catalog subset at the GWOSC \cite{gwtc-1-confident}, and in particular evaluate how well the DAE trained with the O1 LIGO Livingston data only performs on O1 and O2 data gathered by both the LIGO Livingston and LIGO Hanford instruments, see the Gravitational Wave Open Science Center for detailed information \cite{2021SoftX..1300658A}. For brevity, we denote the Livingston detector by L1 and the Hanford detector by H1. The data have been whitened and high-pass filtered with a $f_{low}$ threshold of 30 Hz, as described for the training and testing dataset in Sec. \ref{sect:training_data}. The events' waveforms are taken from the data behind Fig. 10 of the GWTC-1 catalog paper \cite{2019PhRvX...9c1040A}, more precisely from the \emph{lalinference} folder of \cite{link_Fig10}. The plots in \ref{sect:appendixO1} and \ref{sect:appendixO2} display the aleatoric uncertainty estimate discussed in Sec. \ref{sect:aleatoric_uncertainty}.  




\subsection{Model size and training time} 
\label{sect:model_parameters} 

All key parameters of the network (batch size, learning rate, number of units per layer, activation functions) have been determined following a simple grid-based approach. In total, the final version of the DAE model consists of 1491137 parameters (4096 non trainable parameters). The training time on 8000 data instances (7000 GW signals immersed in noise, 1000 glitch instances) is approximately 480 s on NVIDIA Tesla V100-SXM2-32GB, including reading in the training data. The evaluation of the trained DAE takes approximately 20 ms to denoise one time series instance. 

\section{Conclusions}
\label{sect:conclusions}

We introduce a deep-learning technique relying on an AE network architecture to reduce Gaussian and non-Gaussian contributions from BBH GW measurements by ground-based detectors. The model is trained on a population of simulated astrophysical sources injected into real interferometric noise from O1 and O2 LIGO-Virgo observing runs (LIGO LIvingston and Hanford detectors). We studied the efficiency in recovering the SNR and overall waveform from a population of injected signals, and assess the model robustness with respect to some classes of instrumental glitches. Finally, we propose and implement an aleatoric uncertainty estimation method before applying the method to real events observed during the O1 and O2 LIGO-Virgo observing runs (GW events robustly detected and confirmed by other methods). The DAE method presented here is potentially a versatile pre-processing tool prior to detection and/or source parameter estimation pipelines used to analyse data collected by ground based instruments. An immediate step is to extend the source parameter ranges such as the individual masses or the sky localization, and apply the conditional parameter training, as in the CVAE. The approach can also be improved by 

\begin{enumerate}
    \item considering other morphologies of instrumental glitches \cite{2017CQGra..34f4003Z} and increase the training/testing dataset size.
    \item make the loss function more complex, e.g. by adding a regularization term to penalize glitches and unwanted features \cite{2021PhRvD.104f4046C}.
    \item include more ground based detectors in the CNN architecture and see whether it improves noise removal. Note that we have not used the currently-available Virgo data in the analysis because of the low SNR of recovered events.
    \item consider more sophisticated hyperparameter tuning of the network.
    \item improve the uncertainties estimation by e.g. the analysis of the latent space features. 
\end{enumerate}

In summary, we emphasize that we aimed at obtaining a denoising reconstruction method of realistic GW BBH signals with a relatively small training data set (of the order of thousands of samples), and a minimal-size NN. Vast majority of test sample GW signals are recovered sufficiently well, especially the high-frequency part which is also correctly resolved in phase, making the DAE method a potentially computationally-inexpensive trigger generator working in low-latency (a pre-processing step before more expensive parameter estimation methods); the shortcomings presented on e.g. real data examples are generally related to parameters of the GW signals being outside the training dataset parameters, and therefore possible to relieve with training tuned to specific signal parameters (taking into account known techniques to prevent catastrophic forgetting \cite{doi:10.1073/pnas.1611835114}). Improvements in the uncertainties estimation by the DAE model would additionally strengthen its role as a pre-processing step in the general parameter estimation. 

{\small 
{\it Acknowledgements.} The authors would like to acknowledge the European COST action G2Net (CA17137), Tom Charnock and Florian F{\"u}hrer for fruitful discussions, and {\'E}ric Chassande-Mottin for his contribution on the dataset preparation. This research was supported in part by the PLGrid infrastructure with the computing grant on the ACK Cyfronet AGH \texttt{Prometheus} cluster, the grant of the Polish Ministry of Science and Higher Education (MNiST) for the expansion of the IT infrastructure at the Nicolaus Copernicus Astronomical Center, and the National Science Centre grant no. 2016/22/E/ST9/00037. 

Research presented here has made use of data or software obtained from the Gravitational Wave Open Science Center (gw-openscience.org), a service of LIGO Laboratory, the LIGO Scientific Collaboration, the Virgo Collaboration, and KAGRA. LIGO Laboratory and Advanced LIGO are funded by the United States National Science Foundation (NSF) as well as the Science and Technology Facilities Council (STFC) of the United Kingdom, the Max-Planck-Society (MPS), and the State of Niedersachsen/Germany for support of the construction of Advanced LIGO and construction and operation of the GEO600 detector. Additional support for Advanced LIGO was provided by the Australian Research Council. Virgo is funded, through the European Gravitational Observatory (EGO), by the French Centre National de Recherche Scientifique (CNRS), the Italian Istituto Nazionale di Fisica Nucleare (INFN) and the Dutch Nikhef, with contributions by institutions from Belgium, Germany, Greece, Hungary, Ireland, Japan, Monaco, Poland, Portugal, Spain. The construction and operation of KAGRA are funded by Ministry of Education, Culture, Sports, Science and Technology (MEXT), and Japan Society for the Promotion of Science (JSPS), National Research Foundation (NRF) and Ministry of Science and ICT (MSIT) in Korea, Academia Sinica (AS) and the Ministry of Science and Technology (MoST) in Taiwan. 

This material is based upon work supported by NSF's LIGO Laboratory which is a major facility fully funded by the National Science Foundation. 

The code and data were prepared in {\tt python v3.8}, using {\tt Tensorflow/Keras v2.2} with the support of GPU ({\tt CUDA} toolkit v10.1), and the GW libraries {\tt gwpy} v2.0.2 \cite{gwpy} and {\tt pyCBC} v1.17 \cite{2020zndo...4355793N}. We acknowledge the use of {\tt matplotlib} v3.3.3 \cite{Hunter:2007}.

} 

\section{References}
{\small  
\bibliography{paper_dcvae}
} 

\newpage
\appendix

\section{Real O1 GW events}
\label{sect:appendixO1}

In all the plots, the L1 and H1 data are depicted with a blue line. The green line represents the whitened waveform recovered by lalinference (ML waveform) and taken from \cite{link_Fig10}. The red line is the result of the denoising and the yellow band represents the aleatoric uncertainty calculation. The caption of each plot recalls the estimated masses and the single-detector optimal SNRs from parameter-estimation analyses made by the LVC Collaboration in \cite{2019PhRvX...9c1040A}. Below we summarize the reconstructed events, and discuss the reasons of some cases of unsatisfactory reconstructions. 

For the \textrm{GW150914} event in Fig.~\ref{fig:GW150914}, both the amplitude and phase are very well reconstructed in L1 and this despite a clear non-Gaussian contribution at the time of the event. In H1, the phase is well reconstructed while there is a loss of amplitude in the part of the waveform just before the merger.

For the \textrm{GW151012} event in Fig.~\ref{fig:GW151012}, the poor reconstruction may be understood by the magnitude of the single detector optimal SNR found in the parameter-estimation analyses: $6.4^{+1.3}_{-1.3}$ for H1 and $5.8^{+1.2}_{-1.2}$ for L1 \cite{2019PhRvX...9c1040A}. By consulting these values with Fig. \ref{fig:mse_ove_vs_snr} we conclude that the event lies in an area where the overlap is usually not sufficient for satisfactory denoising.    

For the \textrm{GW151226} event in Fig.~\ref{fig:GW151226}, the phase of the signal is well recovered in many parts of the waveform, while the amplitude is almost everywhere underestimated, except for the first and last portion of the event in H1, for which the reconstruction worked. We associate the relatively poor quality of this result with the lighter mass of one of the BH system components, in comparison to the values used in the training. Component mass estimates are $13.7^{+8.8}_{-3.2}$ and $7.7^{+2.2}_{-2.5} \,M_\odot$ \cite{2019PhRvX...9c1040A}, whereas our training set lower mass was $10\,M_\odot$. Lower component masses make the GW signal last longer in time in the sensitive frequency range of the detectors; the DAE failed to generalize to a duration of the GW signal longer than exposed to during the training.

\def\Oonedata{
GW150914/$35.6^{+4.7}_{-3.1}$/$30.6^{+3.0}_{-4.4}$/$14.2^{+1.6}_{-1.4}$/$20.6^{+1.6}_{-1.6}$, 
GW151012/$23.2^{+14.9}_{-5.5}$/$13.6^{+4.1}_{-4.8}$/$5.8^{+1.2}_{-1.2}$/$6.4^{+1.3}_{-1.3}$, 
GW151226/$13.7^{+8.8}_{-3.2}$/$7.7^{+2.2}_{-2.5}$/$6.9^{+1.2}_{-1.1}$/$9.8^{+1.5}_{-1.4}$
} 
\foreach \x\y\z\v\w in \Oonedata
{ 
    \begin{figure}[!ht]
    \begin{center}
        \foreach \d in {L1, H1}
        {
            \IfStrEq{\d}{H1}{\newcommand\trim{2cm}}{\newcommand\trim{1cm}}
            \vspace{5pt} 
            \includegraphics[width=0.85\linewidth, 
            trim={1.5cm 0.5cm 3cm \trim},clip]{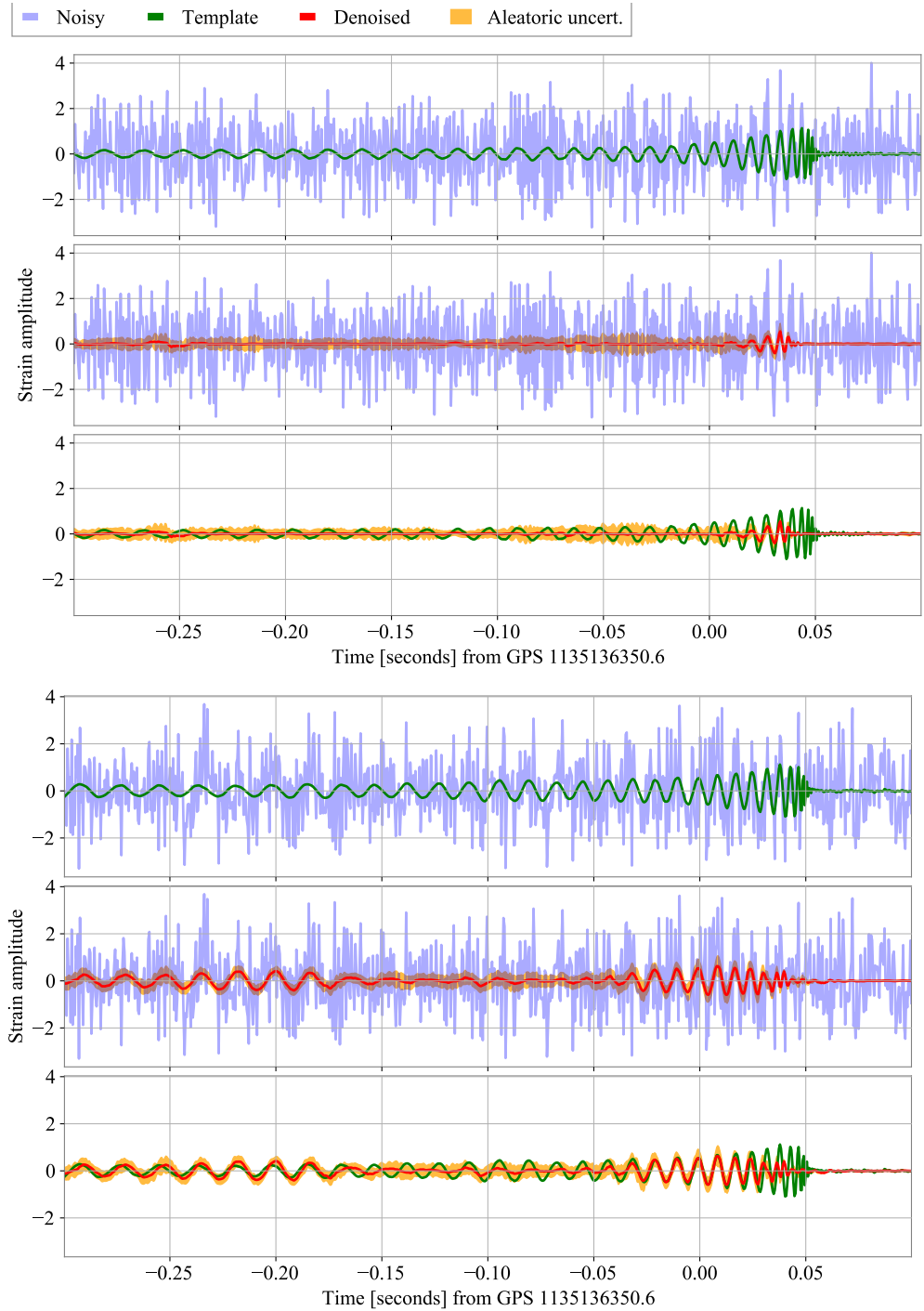}
        }
        \vspace{-5pt}
        \caption{Denoising applied to the O1 data {\x} event for the L1 detector (3 top panels) and the H1 detector (3 bottom panels). The component masses are {\y} $M_\odot$ and {\z} $M_\odot$, and the single-detector optimal SNRs are {\v} for L1 and {\w} for H1.} 
		\label{fig:\x} 
    \end{center}
    \end{figure}
\clearpage        
}


\section{Real O2 GW events}
\label{sect:appendixO2}

The DAE model was also tested on real events from the O2 run; we show selected interesting results in figures analogous to the \ref{sect:appendixO1}, which are mostly cases of non-optimal denoising. As before, we discuss reasons behind these results.
Despite the DAE model was trained on the O1 L1 data only, in case of the louder O2 events whose masses are in the range used for the training, namely the GW170104, GW170809, GW180814, GW170818 events, the waveforms are reconstructed.  

For the \textrm{GW170104} event, Fig.~\ref{fig:GW170104} displays such an example of a clean denoising in both L1 and H1 detectors. Several ($\sim 10$ cycles) high frequency cycles are well-recovered. In general, this event is a good example of the low-frequency part of the signal having too small amplitude and therefore too small SNR to be correctly denoised.  

For the \textrm{GW170823} event in Fig.~\ref{fig:GW170823}, the phase of the original GW signal is well-recovered while an additional ``ghost'' GW signal is inferred from our DAE model (in between $- 0.10$ s and $-0.05$ s) in H1. Note that these spurious cycles do not belong to the original GW signal, i.e. it is not that the clean signal cycles are somewhere well and poorly recovered as it is the case in the previous figure. In this case, the low single detector SNR of this event in H1 (SNR is ${6.8}^{+1.4}_{-1.2}$ in Handford - see Table V in \cite{2019PhRvX...9c1040A}) explains the poor reconstruction. This is coherent with results presented on Fig.~\ref{fig:snri_snrd_snrdno}. We interpret the ``ghost'' signal reconstruction as false-positive reaction of the DAE model on a possible non-Gaussian transient feature in the real data. The issue of deceiving the trained model with specifically-crafted input data is beyond the scope of the current work, but it is a possible future development direction.

For the \textrm{GW170608} event in Fig.~\ref{fig:GW170608}, the DAE model manages to retrieve some cycles in the L1 data, although phase information is not recovered completely. A clearly visible low-frequency glitch in the H1 data prevents the GW signal to be correctly denoised. However, in Fig. \ref{fig:GW170608_50Hz} we perform an experiment with changing the low-pass filter value $f_{low}$, from 30 Hz assumed at training to 50 Hz. In this case the DAE model performs relatively well, although it was trained on the $f_{low}=30$ Hz data, and despite the fact that the source mass parameters lie outside the training dataset parameter space used for training.

\def\Otwodata{
GW170104/$30.8^{+7.3}_{-5.6}$/$20.0^{+4.9}_{-4.6}$/$9.9^{+1.5}_{-1.3}$/$9.5^{+1.3}_{-1.6}$, 
GW170823/$39.5^{+11.2}_{-6.7}$/$29.0^{+6.7}_{-7.8}$/$9.2^{+1.7}_{-1.5}$/$6.8^{+1.4}_{-1.2}$,
GW170608/$11.0^{+5.5}_{-1.7}$/$7.6^{+1.4}_{-2.2}$/$9.2^{+1.5}_{-1.2}$/$12.1^{+1.6}_{-1.6}$
} 

\foreach \x\y\z\v\w in \Otwodata
{ 
    \begin{figure}[!ht]
        \begin{center}
        \foreach \d in {L1, H1}
        {
            \IfStrEq{\d}{H1}{\newcommand\trim{2cm}}{\newcommand\trim{1cm}}
            \vspace{5pt} 
		    \includegraphics[width=0.85\linewidth, trim={1.5cm 0.5cm 3cm \trim},clip]{imgs/test_real_\x_\d_30.0kHz.pdf}
        }
        \vspace{-5pt}
        \caption{Denoising applied to the O2 data {\x} event for the L1 detector (3 top panels) and the H1 detector (3 bottom panels). The component masses are {\y} $M_\odot$ and {\z} $M_\odot$, and the single-detector optimal SNRs are {\v} for L1 and {\w} for H1.} 

		\label{fig:\x}
	    \end{center}
    \end{figure}

    \clearpage
}

\begin{figure}[!ht]
        \begin{center}
        \foreach \d in {L1, H1}
        {
            \IfStrEq{\d}{H1}{\newcommand\trim{2cm}}{\newcommand\trim{1cm}}
            \vspace{5pt} 
		    \includegraphics[width=0.85\linewidth, trim={1.5cm 0.5cm 3cm \trim},clip]{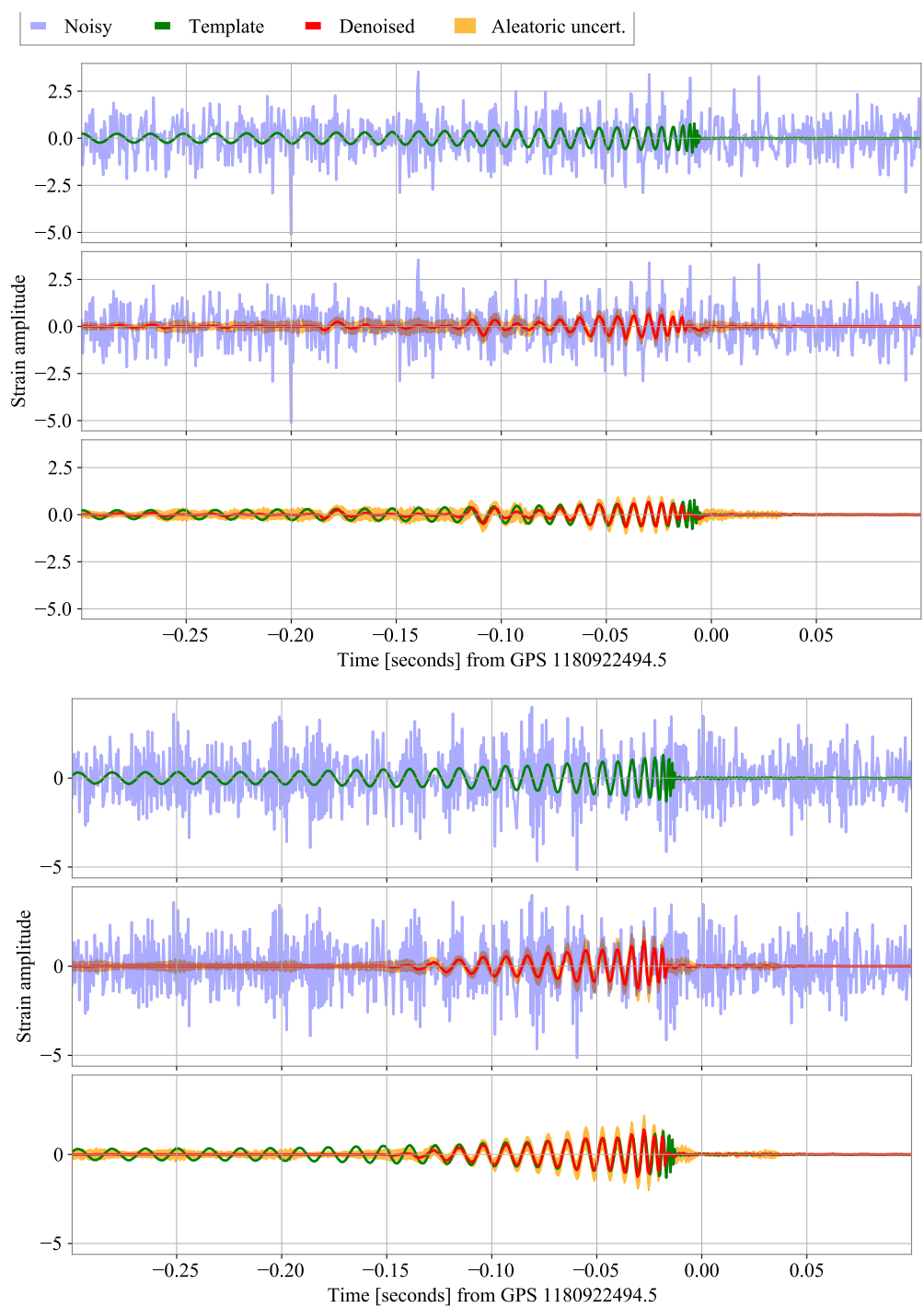}
        }
        \vspace{-5pt}
        \caption{Denoising applied to the O2 data GW170608 event for the L1 detector (3 top panels) and the H1 detector (3 bottom panels). While for the other plots the high-pass filter was set to 30 Hz (as in the training set), in this case we apply high pass at 50 Hz to the original data before the denoising.} 

		\label{fig:GW170608_50Hz}
	    \end{center}
\end{figure}

    \clearpage

\end{document}